\documentclass[preprint,aps,prd,superscriptaddress,nofootinbib,showpacs]{revtex4-1}
\usepackage{graphicx}
\usepackage{epsfig}
\usepackage{bm}
\usepackage{amssymb}
\usepackage{wrapfig}
\usepackage[usenames, dvipsnames]{color}
\usepackage[normalem]{ulem}
\usepackage{color}

\newcommand{\el}{{\cal L}}
\newcommand{\cM}{{\cal M}}
\newcommand{\cO}{{\cal O}}

\newcommand{\epsi}{\mbox{$\varepsilon$}}

\newcommand{\vq}{\mbox{$\bm{q}$}}
\newcommand{\vbr}{\mbox{$\bm{r}$}}

\newcommand{\vB}{\mbox{$\bm{B}$}}

\begin{document}

\title{\bf Axion Production from Landau Quantization \\
in the Strong Magnetic Field of Magnetars
}
\author{Tomoyuki~Maruyama}
\affiliation{College of Bioresource Sciences,
Nihon University,
Fujisawa 252-8510, Japan}
\affiliation{National Astronomical Observatory of Japan, 2-21-1 Osawa,
Mitaka, Tokyo 181-8588, Japan}

\author{A.~Baha~Balantekin}
\affiliation{Department of Physics, University of Wisconsin, Madison,
WI 53706, USA}
\affiliation{National Astronomical Observatory of Japan, 2-21-1 Osawa,
Mitaka, Tokyo 181-8588, Japan}

\author{Myung-Ki~Cheoun}
\affiliation{Department of Physics and Origin of Matter and Evolution of Galaxies (OMEG) Institute, Soongsil University, Seoul,
156-743, Korea}
\affiliation{National Astronomical Observatory of Japan, 2-21-1 Osawa,
Mitaka, Tokyo 181-8588, Japan}

\author{Toshitaka~Kajino}
\affiliation{National Astronomical Observatory of Japan, 2-21-1 Osawa,
Mitaka, Tokyo 181-8588, Japan}
\affiliation{Department of Astronomy, Graduate School of Science,
University of Tokyo, Hongo 7-3-1, Bunkyo-ku, Tokyo 113-0033, Japan}
\affiliation{Beihang University, School of Physics and Nuclear Energy
Engineering, Int. Center for Big-Bang Cosmology and Element Genesis, Beijing 100191, China}
\author{Grant J. Mathews}
\affiliation{Center of Astrophysics, Department of Physics,
University of Notre Dame, Notre Dame, IN 46556, USA}

\date{\today}


\begin{abstract}
 We utilize  
 an exact quantum calculation to  explore axion emission from electrons and protons
in the presence of  the strong magnetic field of magnetars.  
The axion is emitted via transitions between the Landau levels generated
by the strong magnetic field.
The luminosity of axions emitted by protons is shown to be much larger 
than that of  electrons and becomes stronger with increasing matter density.
Cooling by axion emission is shown to be much larger than  neutrino
 cooling by the Urca processes. 
Consequently,  axion emission in the crust may significantly contribute
to the  cooling of magnetars.  In the high-density core, however, 
it may cause heating of the magnetar.
\end{abstract}

\maketitle



The axion is a  hypothetical pseudoscalar particle.   It
is a pseudo-Goldstone boson associated with
the Peccei-Quinn symmetry \cite{PQ77} and has been introduced
as a solution to the strong CP-violation  problem \cite{CP-b,Kim79,DFS81}.
The physics related to the axion has been  discussed in many papers, e.g.
\cite{SVZ80,DFS81,Kim87,Cheng88}.

In particular, axion phenomenology in  astrophysical environments has been extensively explored 
in Refs. \cite{VZKC78,Turner90,Raffelt90,Raffelt96,Raffelt99,Raffelt08}.
Axions are candidates for the cold dark matter of the
universe because they have non-zero mass and their interactions with normal matter should be small.
In view of the lack of detections in recent WIMP searches, the study of axion production or detection is well motivated and axions become a  compelling 
candidate for cold dark matter \cite{AP83,PWW83,DF83,IS83}.
Axion dark matter can  couple to two photons that can subsequently be  observed \cite{MPT16}.
However, various astronomical phenomena and laboratory experimental data \cite{AP83,DFS81,Leison14} have only placed upper limits  on  the axion mass and decay constants.  
Specifically, for hadronic axions the mass and couplings are expected to be 
proportional to each other.

Axions produced in a hot astrophysical plasma can transport energy out of
stars or even reheat the interior plasma if they have a small mean free path.
The strength of the axion coupling with normal matter and radiation
is bounded by the condition that
stellar evolution lifetimes and/or energy loss rates should not conflict with observation. 
Such arguments
can also be applied to the physics of supernova explosions, where the
dominant energy loss processes are thought to be the emission of neutrinos
and anti-neutrinos along with axions via the  mechanism of  
nucleon bremsstrahlung \cite{BT88,BTB89,RS95,HPS01}.

Axions may be efficiently produced in the interiors of stars and act
as an additional sink of energy. Therefore, they can alter
the energetics of some processes, for example, type-II
supernova explosions. Several authors have noted that the
emission of axions ($a$) via the nucleon ($N$) bremsstrahlung process
$N + N \rightarrow N + N + a$ may drain too much energy from type-II supernovae, making them inconsistent
with the observed kinetic energy of such events \cite{BT88,BTB89,BTB90,JKRS96,HPS01}.

In Refs.~\cite{Iwamoto84,UITQN98} the thermal
evolution of a cooling neutron star was studied by including  axion
emission in addition to neutrino energy losses.
An upper limit on the axion mass of 
$m_a < 0.06 -0.3$~eV was deduced. 
Axion cooling is an interesting possibility for
the cooling mechanism of the neutron stars
\cite{Iwamoto84,NKI87,NAKI88,Weber98,Iwamoto01,Sedrakin07,Sedrakin16}.
In their pioneering study, Umeda et al. \cite{UITQN98} considered the  axion
radiation produced  via the bremsstrahlung in $NN$ collisions
in bulk nuclear matter. Axion emission from a meson condensate
\cite{MTI94} was also studied.

Cosmological constraints may also provide upper and lower limits
on the mass of the axion \cite{Rosenb15}.  
Nevertheless, there still remains a large region
of the parameter space to be searched.
One of the most well developed and sensitive
experiments is the Sikivie haloscope \cite{Sikivie83,Sikivie85}.  
This approach exploits the inverse Primakoff effect whereby a magnetic field
provides a source of virtual photons in order to induce axion-to-photon
conversion via a two-photon coupling.
The generated real photon frequency is then determined  by the axion mass.
This signal can be  resonantly enhanced by a cavity structure and
resolved above the thermal noise of the measurement system.
It has been proposed  \cite{Sikivie83,Sikivie85,HHMSSTH11} that in a haloscope with an axial DC magnetic
field the expected power due to axion-to-photon conversion
can be detected.

The present status of the mass and coupling constant are well summarized and tabulated in Ref. \cite{PDG16}.
Lower limits exist for  the coupling constant, $g_{a\gamma \gamma}$,
in the Lagrangian,
\begin{equation}
\el_{a \gamma \gamma} = 
- {g_{a \gamma \gamma} \over 4} F_{\mu \nu} {\tilde F}^{\mu \nu} \phi_A ~,
\end{equation}
where $\phi_A$ is the axion field and $F_{\mu \nu}$ is the
electro-magnetic field strength tensor.  
Currently, from Helioscopes, $|g_{a\gamma \gamma}| < 
6.6 \times 10^{-11}$~GeV$^{-1}$ (95 \% CL) for a mass range of, 
$10^{-10}~{\rm eV} < m_a < 1$~eV.
In addition, the analysis of gamma-rays from  SN1987A  \cite{PEFGMR15} has led to the constraint that  
$|g_{a \gamma \gamma}| \lesssim 5.3 \times 10^{-12}$~GeV$^{-1}$ 
and $m_a < 4.4 \times 10^{-10}$~eV.

Axion couplings for fermions, $g_{aNN}$ and $g_{aee}$, in the Lagrangian
\begin{equation}
{\cal L}_{a f f} = - i g_{a f f} 
{\tilde \Psi}_f \gamma^{\mu} \gamma_5 \Psi_f \phi_a
\end{equation}
 are constrained to be  $\alpha_{aee}  = g_{aee}^2 / 4 \pi < 1.5 \times 10^{-26}$
and $g_{a NN} = (3.8 \pm 3) \times 10^{-10}$
based upon  many experiments and observations \cite{PDG16}.

\bigskip

On the other hand, magnetic fields in neutron stars are much  stronger
than those in laboratory experiments,
Hence, axion emission may   play a vital role in the interpretation
of many observed phenomena.
In particular, magnetars, which are associated with super-strong magnetic fields,
\cite{pac92,mag3} have many exotic features that distinguish them  from
normal the neutron stars. Thus, phenomena associated  with  magnetars
can give information about the physical processes associated with  strong magnetic fields.

It has been noted  \cite{nakano11} that the characteristic magnetar spin down periods ($P/2 \dot P$) 
(where $P$ is the spin period)
appear to be systematically overestimated compared to
the ages of the associated supernova remnants.
Soft gamma repeaters (SGRs) and anomalous X-ray pulsars (AXPs)
are believed to be  to magnetars \cite{Mereghetti08}. 
Magnetars emit energetic photons.
Furthermore, the surface temperature of the magnetars is 
$T\approx 280  - 720~$eV.
This is larger than that  of normal neutron star which typically have a surface temperature of $T \approx 10 - 150~$eV
for similar ages \cite{Kaminker09}.
Thus, the associated strong magnetic fields may have
significant effects on these objects, and 
there must be a mechanism to convert the magnetic energy into
thermal and radiant energies.

In this work we calculate the axion emission due to electrons and protons
in the Landau quantization of the strong magnetic field.  This
 mechanism  is different from the previously considered bremsstrahlung or
Primakoff mechanisms for axion production.
Such axion emission from electrons has been previously  calculated classically and
quantum mechanically \cite{Bori94,KWW97}. 
However, the emission from protons was not taken into account.
Here we show that the axion luminosity expected from the protons 
inside a magnetar is much larger than that due to electrons and it is high 
enough to be considered in the
neutron star cooling (or reheating) process.
In particular, contributions from the anomalous magnetic moment (AMM) of the 
protons becomes significant, as has been discussed in the pion production by the
magnetic field \cite{P2Pi-1,P2Pi-2}.


\bigskip


We assume a uniform magnetic field along the $z$-direction,
$\vB = (0,0,B)$, and take the electro-magnetic vector potential $A^{\mu}$ to be
$A = (0, 0, x B, 0)$ at the position $\vbr \equiv (x, y, z)$. 
The relativistic wave function $\psi$ is obtained
from the following Dirac equation:
\begin{equation}
\left[ \gamma_\mu \cdot (i \partial^\mu - \zeta e A^\mu - U_0 \delta^0_\mu)
- M + U_s
- \frac{e \kappa}{2 M} \sigma_{\mu \nu}
(\partial^\mu A^\nu - \partial^\nu A^\mu ) \right]
\psi_a (x) = 0 ,
\label{DirEq}
\end{equation}
where $\kappa$ is the AMM, 
$e$ is the elementary charge
and $\zeta =\pm 1$ is the sign of the particle charge.
$U_s$ and $U_0$ are the scalar field and  time component of the vector
field, respectively.

In our model charged particles are protons and electrons.   
The mean-fields are taken to be zero for electrons, while 
for protons they are given by  relativistic mean-field
(RMF) theory~\cite{serot97}.
The single particle energy is then written as
\begin{eqnarray}
&& E(n, p_z, s) = \sqrt{ p_z^2 + (\sqrt{2 eB n + M^{*2}}
- s e  \kappa B /M)^2} + U_0
\label{Esig}
\end{eqnarray}
with $M^* = M - U_s$,
where $n$ is the Landau number, 
$p_z$ is a $z$-component of momentum, and $s = \pm 1$ is the spin.
The vector-field $U_0$ plays the role of shifting  the single particle energy
and does not contribute to the result of the calculation.  Hence,  
we can omit the vector field in what  follows.

We obtain the differential decay width of the proton from  the pseudo-vector coupling for the axion-proton (electron)
interaction, 
\begin{equation}
\frac{d^3 \Gamma}{d \vq^3} =
\frac{g_a^2}{8 \pi^2 e_a}
\sum_{n_f,s_f}  \frac{\delta(E_f + e_a - E_{i})}{4 E_i E_f} W_{if}~
f(E_i) \left[ 1 - f(E_f) \right] ~~,
\label{dfWid}
\end{equation}
with
\begin{equation}
W_{if} = {\rm Tr} \left\{ \rho_M (n_i, s_i, p_z) \cO_{A}
\rho_M (n_f, s_f, p_z - q_z) \cO_{A}^{\dagger}
 \right\} ,
\end{equation}
where $e_a$ is the energy of the emitted axion,
$\vq \equiv (q_x, q_y, q_z)$ is the axion momentum,
$g_a$ is the pseudo-scalar axion coupling constant,
and
\begin{eqnarray}
\rho_M &=&
\left[ E \gamma_0 + \sqrt{2 eB n} \gamma^2 - p_z \gamma^3
 + M^* + (eB \kappa/M) \Sigma_z \right]
 \nonumber \\ && \quad \times
\left[ 1 + \frac{s}{\sqrt{ 2 eB n + M^{*2}} } \left(
 eB \kappa /M + p_z \gamma_5 \gamma_0 + E \gamma_5 \gamma_3 \right)
 \right] ,
\end{eqnarray}
%
while
\begin{eqnarray}
\cO_{A}
&=&
\gamma_5 \left[
\cM \left( n_i, n_f \right) \frac{1 + \zeta \Sigma_z}{2}
+  \cM \left( n_i - 1, n_f - 1 \right) \frac{1 - \zeta \Sigma_z}{2}
	 \right]~~ .
\end{eqnarray}

In the above equation,
the harmonic oscillator (HO) overlap function $\cM(n_1, n_2)$
is defined as \cite{P2Pi-1, P2Pi-2}
\begin{eqnarray}
 \cM (n_1,n_2)  & =&
\int^{\infty}_{-\infty} d x h_{n_1}\left( x - \frac{q_T}{2 \sqrt{eB}} \right)
h_{n_2} \left( x+ \frac{q_T}{2 \sqrt{eB}} \right) ,
\label{TrStM}
\end{eqnarray}
where $q_T = \sqrt{q_x^2 + q_y^2}$, and $h_n (x)$ is the HO wave
function with  quantum number $n$.

The mass and coupling constants of the axion are still ambiguous.
The axion mass is much smaller than the energy difference between
 different Landau levels in the present work, 
and its value does not affect the final results. 
In this work we choose
the axion-nucleon coupling to be $g_{aNN} = 6 \times 10^{-12}$ and
the axion-electron coupling to be $g_{aee} = 9 \times 10^{-15}$,
which are $10^{-2}$ below the maximum value deduced in Ref.~\cite{Iwamoto84}.
These parameters are chosen to impose the condition that the axion
emission be negligible compared to the neutrino emission
in normal neutron stars.

Furthermore, we use the parameter-sets in Ref.~\cite{MHKYKTCRM14}
for the equation of state (EOS) of the neutron-star matter, 
which we take to be comprised of neutrons, protons and electrons.
In this work we take the temperature to be very low, $T \ll 1~$MeV,
and use the mean-fields at zero temperature.


\begin{figure}[bht]
\begin{center}
{\includegraphics[scale=0.6,angle=270]{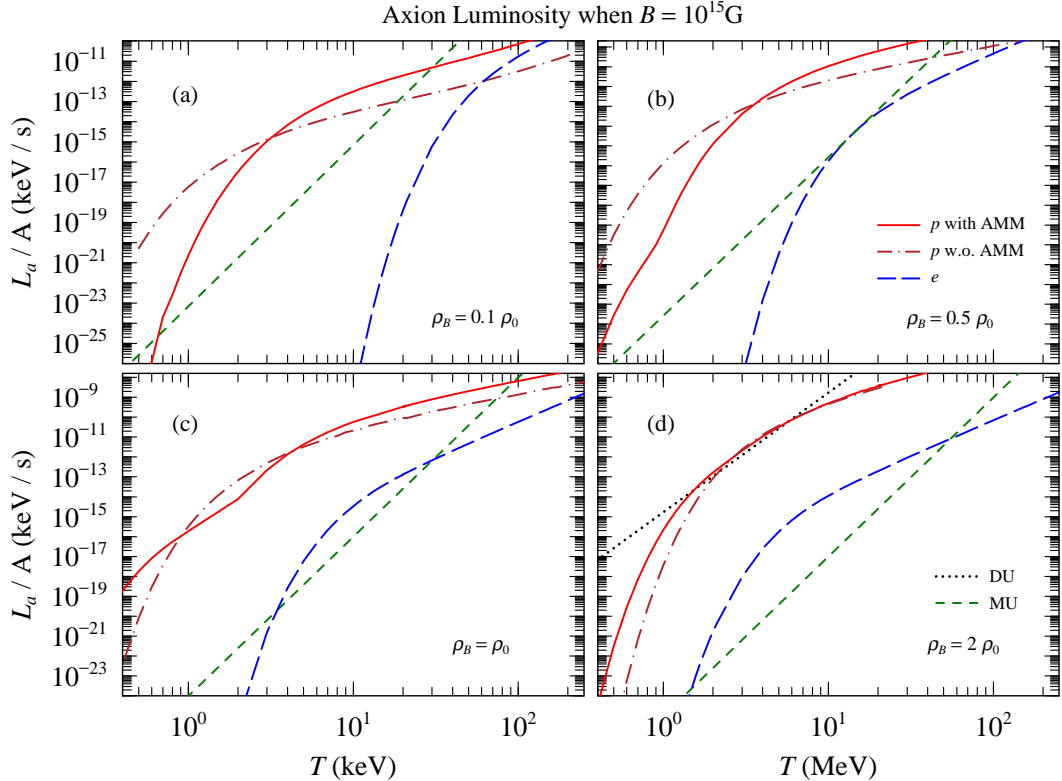}}
\caption{\small
(Color online) Axion luminosity versus temperature at baryon densities 
$\rho_B = 0.1 \rho_0$ (a) $\rho_B = 0.5 \rho_0$ (b),
$\rho_B = \rho_0$ (c) and $\rho_B = 2 \rho_0$ (d) for $B = 10^{15}$G.
The solid and dot-dashed lines  represent the results for protons with and
 without the AMM, respectively.
The long-dashed line indicates the results for electrons.
The dashed and dotted lines indicate the neutrino luminosities from 
the MU and DU processes, respectively.
}
\label{L-Tdep}
\end{center}
\end{figure}

In Fig.~\ref{L-Tdep} we show the temperature dependence of the axion
luminosity per nucleon at $B=10^{15}$G for baryon densities of: (a) $\rho_B = 0.1 \rho_0$; (b)   $\rho_B = 0.5 \rho_0$; (c) 
$\rho_B = \rho_0$;  and  (d) $\rho_B = 2 \rho_0$.
The solid, dot-dashed and long-dashed lines represent the contributions
from  protons with the AMM,  without the AMM, and that of electrons,
respectively.
For comparison, we also exhibit the neutrino luminosities from
the modified Urca (MU) process \cite{MURCA} (dashed lines)  and
those from the direct Urca (DU) process \cite{DURCA} (dotted lines).
(Note that the contribution from the AMM is omitted in the DU process,)

First, we see that the axion luminosity varies slowly when $T
\gtrsim 10$ keV,  while it changes rapidly in the low temperature region.
It is well known that the low temperature expansion leads to a power law  temperature dependence of the 
 the emission luminosity, {\it i.e.}, ${\cal L} = c T^a$.

In the semi-classical approach \cite{BG-JETP94},
the axion luminosity from an electron was shown to be
proportional to $T^a$ with $a = 13/3  \approx 4.3$.
In our results the electron contributions can be fitted
with $a = 3.6-3.8$ in the high temperature region; 
these values are similar to those obtained in the semi-classical approach.
However, one should also consider realistic low magnetar  temperatures
$T \lesssim 1~$keV.  
In this case, the temperature dependence of the luminosity is more complicated.
In particular, to satisfy the power law, one  requires that the particle 
energies be  continuous. In a strong magnetic field, however, 
the transverse momentum is discontinuous.

The  energy of the emitted axion, $\epsi_a$, for a charged particle transition is obtained as
\begin{eqnarray}
\epsi_a &=& E (n_i, p_z, s_i) - E (n_f, p_z - q_z, s_f)
\nonumber \\
&=& \sqrt{2 eB n_i + p_z^2 + M^{*2}}
- \sqrt{2 eB (n_i - \Delta n_{if}) + (p_z - q_z)^2 + M^{*2}}
- \frac{eB \kappa}{M} \Delta s_{if}
\nonumber \\ &\approx&
\frac{eB}{\sqrt{2 n_i eB + M^{*2}}}  \Delta n_{if}
+ \frac{p_z q_z}{\sqrt{2 n_i eB + M^{*2}}}
- \frac{e B \kappa}{M} \Delta s_{if} ,
\end{eqnarray}
where $\Delta n_{if} = n_i - n_f$,  $\Delta s_{if} = (s_i - s_f)/2$,
and $n_{i,f} \gg \Delta n_{if}$ is assumed.

As the initial Landau number increases, the decay width for PS particle
emission becomes larger \cite{P2Pi-2}, and the state at
$p_z \approx q_z \approx 0$ gives
the largest contribution.
Furthermore,
the energy of emitted particles at the largest decay strength
is proportional to the  mass of the produced particle \cite{P2Pi-2}.
The axion mass is negligibly small, and the largest contribution comes from
$\Delta n_{if} = 1$.

In the low temperature region,  the initial and final states are near
the Fermi surface and $p_z \approx q_z \approx 0$,
so that the energy interval of the dominant transition
is given by
\begin{equation}
e_a \approx \Delta E = \frac{eB}{E_F^*} - \frac{e B \kappa}{M} \Delta s_{if} .
\end{equation}
with $E_F^* = E_F - U_0$, where $E_F$ is the Fermi energy.

The luminosities are proportional to the Fermi distribution of
the initial state and the Pauli-blocking factor of the final state, $f(E_i) [1-f(E_f)]$.
In the low temperature expansion, it is assumed that
energies of the initial and final states populate  the region with
$E_F - T \lesssim E_{i,f} \lesssim E_F + T$ because
the factor $f(E_i)[1 -f (E_f)]$ becomes very small except in this region.

When $T \lesssim \Delta E \approx eB/E_F^*$,
neither the initial nor the final states reside in the above region.
Hence, the luminosities rapidly decrease at low temperature 
as the magnetic field becomes weaker.

When  $B=10^{15}$G,  $\sqrt{eB} = 2.43$~MeV, $eB /E_F^* = 6.6$~keV at $\rho_B = 0.1 \rho_0$,
while  $eB /E_F^* = 9.4$~keV at  $\rho_B =\rho_0$  for protons,  and
$eB /E_F^* = 43$~keV at $\rho_B = 0.1 \rho_0$.  For electrons 
 $eB /E_F^* = 6.7$~keV at  $\rho_B =\rho_0$.
As can be seen in  Fig.~\ref{L-Tdep}, indeed, the change of the axion luminosities
becomes  more abrupt for $T \lesssim eB/E_F^*$.

The energy step is much larger for protons than
electrons because the proton mass is much larger than the electron mass,
and the proton axion luminosity becomes the dominant source.

Furthermore, one can see that there are  shoulders in the density
dependence of the luminosity for protons with the AMM included 
at $T \sim 1$ keV when $\rho_B = 0.5 \rho_0$
and at $T \sim 2$ keV when $\rho_B =  \rho_0$.
The transition of $s_i = -1 = - s_f$ is dominant in the higher
temperature region while the transition $s_i = +1 = - s_f$ becomes
dominant in the lower temperature region.
The spin non-flip transition seldom contributes to the emission of 
PS particles \cite{P2Pi-1, P2Pi-2}.
The roles of the two contributions reverse at the temperature
of the shoulders.
In addition, this reversal occurs at $T \sim 3~$keV
when $\rho_B = 0.1 \rho_0$
though the shoulder is not very evident.

When $\rho_B =\rho_0$ and $B=10^{15}$G, $eB \kappa/M = 7.43$~keV.
In the transition of $s_i = -1 = -s_f$, the AMM interaction for
the initial state is repulsive, while at the final state attractive. 
The additional energy contributes to the transition.
When the temperature is high enough, this positive additional energy
causes the luminosity to increase.
When the temperature is very low, however, the positive additional energy makes
the energy interval $\Delta E$ larger than the temperature.  This 
 suppresses the luminosity.

\begin{figure}[htb]
\begin{center}
{\includegraphics[scale=0.5,angle=270]{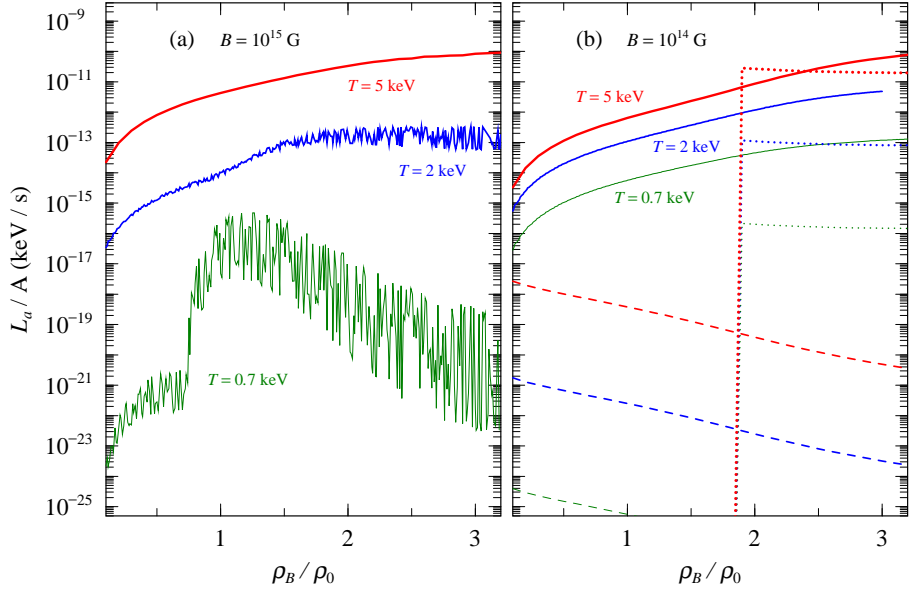}}
\caption{\small
(Color online) Axion luminosity versus baryon density at temperatures
$T =  0.7$~keV ,  $T = 2$~keV and  $T = 5$~keV (from bottom to top))
 when $B = 10^{15}$G (a) and  $B = 10^{14}$G (b).
The dashed and dotted lines in the right panel (b) indicate
the results of the MU and DU processes.
}
\label{TLrrDp}
\end{center}
\end{figure}

In Fig.~\ref{TLrrDp} we show the density dependence of the total axion
luminosity for  $B = 10^{15}$~G (a) and  $B = 10^{14}$~G (b).  
The solid lines show the results  at $T = 0.7$~keV, 2~keV and 5~keV 
from below to above.
For comparison, we plot the neutrino luminosities in the DU process (dotted
line) and those in the MU processes (dashed lines) in the right panel
(b), which are independent of the magnetic field strength.

The luminosity  at $T=0.7$~keV first increases and then
decreases with some fluctuations as the baryon density increases.
All other results  increase monotonously,
but they become more or less saturated at higher densities.

As argued before, the luminosity is mainly determined by the factor
$f(E_i)[1-f(E_f)]$.
The $z$-component of the momentum is not changed much for  the
PS-particle emission \cite{P2Pi-2}, and
$E_i$ and $E_f$ can be thought of as having  discrete energy levels
so that the density dependence of the factor $f(E_i)[1-f(E_f)]$ 
does not smoothly for strong magnetic fields and very low temperatures.

In addition, the axion luminosities are much larger than that of 
neutrinos in the MU process in the present calculation 
even when we take the coupling 
constant to be $10^{-2}$ of the upper limit in Ref.~\cite{Iwamoto84}.
So, the axion luminosity can be expected to give an important
contribution to magnetar cooling.

Furthermore, we notice that the results at $T=0.7$ and 1~keV are
smaller at $B=10^{15}$~G than those for $B= 10^{14}$~G.  This is counter 
intuitive:
the luminosity  becomes larger as the magnetic field increases.  
When $B=10^{14}$~G,  $eB /E_F^* = 0.04$~keV and
$e\kappa_p/M =4.8$~keV for protons, and the discretization of energy levels does
not contribute to the final results.
Indeed, the results for  $T=5$~keV are larger at $B=10^{15}$~G than
that at  $B=10^{14}$~G.

One can attempt to determine the upper limit of  the axion
coupling constant from the calculation results.
One usually expects the axion luminosity to not exceed 
the (anti-)neutrino luminosity in neutron star cooling.
As discussed above, axions produced in a low density region
contribute to the neutron star cooling, 
which is dominantly caused by the MU process.
Then, we use  $4.0 \times 10^{-25}$~keV, which is
the anti-neutrino luminosity for the MU process per nucleon 
at $T=0.7~$keV and $\rho_B = 0.1 \rho_0$, as a baseline value. 

\begin{wrapfigure}{r}{9.5cm}
\begin{center}
{\includegraphics[scale=0.47,angle=270]{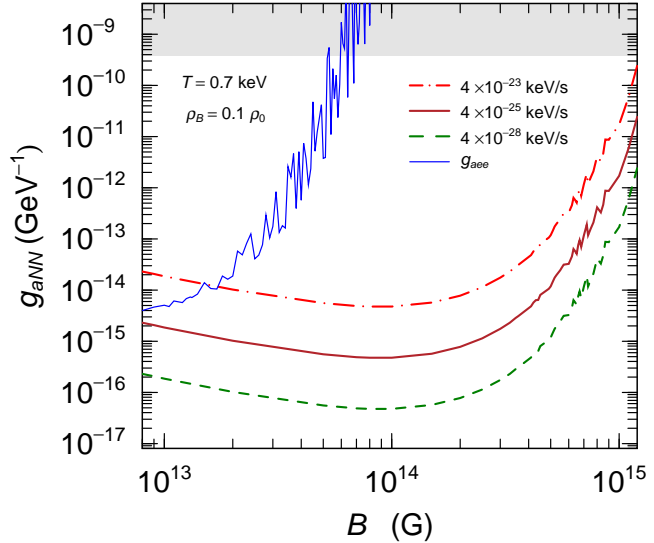}}
\caption{\small
(Color online) The upper limits to the axion-nucleon coupling
at $T = 0.7$~keV and $\rho_B = 0.1 \rho_0$.
The dot-dashed, solid and dashed lines represent results
when the maximum axion luminosities per nucleon are 
$\el_a = 4.0 \times 10^{-23}$~keV/s,  
$4.0 \times 10^{-25}$~keV/s and $4.0 \times 10^{-27}$~keV/s, respectively. 
The dotted lines indicate the electron-axion coupling 
when $\el_a = 4.0 \times 10^{-25}$~keV.
The shade region indicates values ruled out 
by experimental results \cite{PDG16}. }
\label{Gc-Bdep}
\end{center}
\end{wrapfigure}

In Fig.~\ref{Gc-Bdep} we show the magnetic field dependence of the
maximum axion coupling at $T=0.7~$keV and $\rho_B = 0.1 \rho_0$.
The dot-dashed, solid and dashed lines represent the upper limits to  
the axion-nucleon coupling constant $g_{aNN}$  with 
the maximum luminosity being  $4.0 \times 10^{-23}$~keV/s,
 $4.0 \times 10^{-25}$~keV/s and  $4.0 \times 10^{-28}$~keV/s, respectively.

The shaded region exhibits the region 
$g_{aNN} \ge 3.8 \times10^{-10}~$GeV$^{-1}$, which is the present upper limit.
The upper limits of $g_{aNN}$  are much lower than this value.

Furthermore, one can see that $g_{aNN}$ obtains minimum at a minimum values
at $B \approx 9 \times 10^{13}$G. 
This indicates that the luminosity is maximum at this strength of $B$.

In addition, the dotted line indicates the upper limit of  the
axion-electron coupling $g_{aee}$  
when the maximum luminosity is   $4.0 \times 10^{-25}$~keV.
It is shown to increase in an oscillatory manner 
with increasing magnetic field.
So, the strength of the magnetic field which gives the maximum luminosity
is less than $B=10^{13}~$G. 
It was shown in Ref.~\cite{KWW97} that  the axion luminosity 
from electrons decreases with a similar oscillation manner
when $B > m_e^2/e \approx 4.41 \times 10^{13}$~G
at a temperature of $T \ge 5$~keV and an electron density
of $\rho_e = 10^{-4}$~fm$^{-3} \approx 0.006~\rho_0$.

We give a comment on the $B$ dependence of the axion luminosity.

As mentioned above, the discontinuity of the energy levels affects the results
in regions of large  magnetic field strength.
Assuming this to be a generic behavior,  
the peak magnetic field is given by the following equation 
\begin{equation}
T \sim \frac{eB_{max}}{E_F^*}, \qquad
B_{max} \sim T E_F^* /e.
\end{equation}
With $T = 0.7~$keV, we estimate that the magnetic field strength at the maximum is
$B_{max} \sim 8.4 \times 10^{13}~$G for protons  and
$B_{max} \sim 1.7 \times 10^{12}~$G for electrons.
This estimate for protons is close to the exact calculated results.

As the baryon density increases,  the $E_F^*$ of protons 
slightly decreases with $B$ in the density region considered here.  
This is because $B_{max}$ does not have a strong density dependence.
On the other hand, since the $E_F$ for  electrons increases,  $B_{max}$ must
become smaller as the density increases.

It is well known that the axion can couple with two photons, and
that the axion and the real photon are mixed  in a  magnetic field and
oscillate as $a  \rightarrow \gamma \rightarrow a \rightarrow \cdots$
\cite{Raffelt96,Raffelt88}.

In the large magnetic field limit
the wave length is given by $\lambda \approx 4 \pi /|g_{a \gamma \gamma}|B$.
When $|g_{a \gamma \gamma}| \lesssim 5.3 \times 10^{-12}$~GeV$^{-1}$
\cite{PEFGMR15},  $\lambda \gtrsim 120~$fm.
Even if the coupling becomes smaller, the wavelength must be much smaller
than the magnetar radius, $\sim 10$ km.
Many of the photons are absorbed by charged particles, 
so that most of the axions are absorbed by the medium.

We can also ask if axion production could be more
effective in normal neutrons stars in addition to  magnetars.
In the weak magnetic fields of normal neutron stars, 
the energy intervals are very small and calculations become more involved.
We defer this topic to a future publication.
Nevertheless, since our results show that the axion luminosity is much larger
than that due to  neutrinos, in a future publication  we plan to consider
axion emission from normal neutron-stars in the relativistic quantum approach.

In summary, we have studied  axion emission from neutron-star matter
with the strong magnetic fields, $B=10^{15}$G and $10^{14}$~G
in the relativistic quantum approach. We calculated the axion luminosities due to  the transitions of protons and
electrons between two different  Landau levels without invoking any any classical
approximation. 

The axion luminosities turn out to be much larger than that of  neutrinos  due to  the MU
process in the present calculation even when we take the coupling
constant to be $10^{-2}$ of the upper limit.
The axion couplings are not yet completely constrained, but
 our axion luminosity is about $10^{4}$ times  larger than the neutrino
luminosity when $B=10^{15}$~G and $\rho_B = 0.1 \rho_0$. 
Therefore, the axion luminosity can be expected to make  an important
contribution to  magnetar cooling. 
One more point to be noted  is that magnetic fields of about $10^{14}$G 
may be present, leading to a maximum in the axion luminosity at low temperatures. 

Fully quantum calculations provide a higher yield for particle
production  than the semi-classical and/or the
perturbative calculations for pions \cite{P2Pi-1,P2Pi-2} and axions. 
Hence it would be worthwhile to investigate the heating processes 
of magnetars \cite{Heat}
by calculating particle production from other mechanisms 
such as photons from  synchrotron radiation in the quantum approach.

\acknowledgements
This work was supported in part by the Grants-in-Aid for the Scientific
Research from the Ministry of Education, Science and Culture of
Japan~(16K05360), in part by the National Research Foundation of Korea 
(NRF-2014R1A2A2A05003548 and NRF-2015K2A9A1A06046598), 
and in part by the US National Science Foundation Grant No. PHY-1514695. 
Work at the University of Notre Dame is supported
by the U.S. Department of Energy under 
Nuclear Theory Grant DE-FG02-95-ER40934.

\end{document}